\documentstyle[amsfonts,amssymb,twocolumn,eepic]{mn}

\def\bx{\bmath x}
\def\by{\bmath y}
\def\bk{\bmath k}
\def\bq{\bmath q}
\def\bp{\bmath p}
\def\bu{\bmath u}
\def\bw{\bmath w}
\def\bQ{\bmath Q}
\def\pI{Paper~I}

\newif\ifAMStwofonts

\ifoldfss
  \ifCUPmtlplainloaded \else
    \NewTextAlphabet{textbfit} {cmbxti10} {}
    \NewTextAlphabet{textbfss} {cmssbx10} {}
    \NewMathAlphabet{mathbfit} {cmbxti10} {} 
    \NewMathAlphabet{mathbfss} {cmssbx10} {} 
  \fi
  \ifAMStwofonts
    \ifCUPmtlplainloaded \else
      \NewSymbolFont{upmath} {eurm10}
      \NewSymbolFont{AMSa} {msam10}
      \NewMathSymbol{\upi}     {0}{upmath}{19}
      \NewMathSymbol{\umu}     {0}{upmath}{16}
      \NewMathSymbol{\upartial}{0}{upmath}{40}
      \NewMathSymbol{\leqslant}{3}{AMSa}{36}
      \NewMathSymbol{\geqslant}{3}{AMSa}{3E}

      \let\leq=\leqslant 
      \let\geq=\geqslant 
    \fi
  \fi
\fi 

\ifnfssone
  \newmathalphabet{\mathit}
  \addtoversion{normal}{\mathit}{cmr}{m}{it}
  \addtoversion{bold}{\mathit}{cmr}{bx}{it}
  \newmathalphabet{\mathbfit} 
  \addtoversion{normal}{\mathbfit}{cmr}{bx}{it}
  \addtoversion{bold}{\mathbfit}{cmr}{bx}{it}
  \newmathalphabet{\mathbfss} 
  \addtoversion{normal}{\mathbfss}{cmss}{bx}{n}
  \addtoversion{bold}{\mathbfss}{cmss}{bx}{n}
  \ifAMStwofonts
    \ifCUPmtlplainloaded \else
      \UseAMStwoboldmath
      \makeatletter
      \new@mathgroup\upmath@group
      \define@mathgroup\mv@normal\upmath@group{eur}{m}{n}
      \define@mathgroup\mv@bold\upmath@group{eur}{b}{n}
      \edef\UPM{\hexnumber\upmath@group}
      \new@mathgroup\amsa@group
      \define@mathgroup\mv@normal\amsa@group{msa}{m}{n}
      \define@mathgroup\mv@bold\amsa@group{msa}{m}{n}
      \edef\AMSa{\hexnumber\amsa@group}
      \makeatother
      \mathchardef\upi="0\UPM19
      \mathchardef\umu="0\UPM16
      \mathchardef\upartial="0\UPM40
      \mathchardef\leqslant="3\AMSa36
      \mathchardef\geqslant="3\AMSa3E

      \let\leq=\leqslant 
      \let\geq=\geqslant 
    \fi
  \fi
\fi 

\ifnfsstwo
  \DeclareMathAlphabet{\mathbfit}{OT1}{cmr}{bx}{it}
  \SetMathAlphabet\mathbfit{bold}{OT1}{cmr}{bx}{it}
  \DeclareMathAlphabet{\mathbfss}{OT1}{cmss}{bx}{n}
  \SetMathAlphabet\mathbfss{bold}{OT1}{cmss}{bx}{n}
  \ifAMStwofonts
    \ifCUPmtlplainloaded \else
      \DeclareSymbolFont{UPM}{U}{eur}{m}{n}
      \SetSymbolFont{UPM}{bold}{U}{eur}{b}{n}
      \DeclareSymbolFont{AMSa}{U}{msa}{m}{n}
      \DeclareMathSymbol{\upi}{0}{UPM}{"19}
      \DeclareMathSymbol{\umu}{0}{UPM}{"16}
      \DeclareMathSymbol{\upartial}{0}{UPM}{"40}
      \DeclareMathSymbol{\leqslant}{3}{AMSa}{"36}
      \DeclareMathSymbol{\geqslant}{3}{AMSa}{"3E}

      \let\leq=\leqslant 
      \let\geq=\geqslant 
    \fi
  \fi
\fi 

\ifCUPmtlplainloaded \else
  \ifAMStwofonts \else 
    \def\upi{\pi}
    \def\umu{\mu}
    \def\upartial{\partial}
  \fi
\fi

\begin{document}

\title[Study of corrections to the dust model]{Study of corrections to the dust model via perturbation theory}

\author[A. Dom\'\i nguez]{Alvaro Dom\'\i nguez 
\thanks{email: alvaro@fluids.mpi-stuttgart.mpg.de} \\
  Max-Planck-Institut f\"ur Metallforschung, Heisenbergstrasse 1, D-70569 Stuttgart, Germany}

\date{Accepted 2002 March 25. Received 2002 March 15; in original form 2001 June 20}

\maketitle

\volume{334}
\pagerange{435-443}
\pubyear{2002}

\begin{abstract}
  
  This work reports on the application of the Eulerian perturbation
  theory to a recently proposed model of cosmological structure
  formation by gravitational instability \cite{Domi00}. Its physical
  meaning is discussed in detail and put in perspective of previous
  works. The model incorporates in a systematic fashion corrections to
  the popular dust model due to multistreaming and, more generally,
  the small-scale, virialized degrees of freedom. It features a
  time-dependent length scale $L(t)$ estimated to be $L/r_0 \sim
  10^{-1}$ ($r_0 (t)$ is the nonlinear scale, at which $\langle
  \delta^2 \rangle = 1$). The model provides a new angle on the dust
  model and allows to overcome some of its limitations. Thus, the
  scale $L(t)$ works as a physically meaningful short-distance cutoff
  for the divergences appearing in the perturbation expansion of the
  dust model when there is too much initial power on small scales. The
  model also incorporates the generation of vorticity by tidal forces;
  according to the perturbational result, the filtered vorticity for
  standard CDM initial conditions should be significant today only at
  scales below $\sim 1\ h^{-1}$Mpc.

\end{abstract}


\section{Introduction}

The evolution of cosmological structures by gravitational instability
is usually modelled with the popular dust model \cite{Peeb80,SaCo95}:
it is simply the hydrodynamic Eqs.~for a fluid under the influence of
no other force but its own gravity (no pressure, no viscosity, no heat
flows). The usual justification of this model is that (long-range)
gravity is the overwhelmingly dominant force on the cosmological large
scales and that, while the gravitational instability is not too
advanced, the matter distribution looks like a continuum. It is found,
however, that the success of the dust model extends beyond the
expectations raised by this argumentation, as evidenced by the
comparison with N-body simulations of the perturbation theory
predictions, both Eulerian \cite{JBC93,Bern94,GaBa95,Bern96} and
Lagrangian \cite{BMW94,MBW95,BCHJ95,WGB96}: some perturbation
predictions hold even close to the nonlinear scale $r_0$, when the
matter distribution does not look homogeneous any more. But the dust
model has of course its own shortcomings, notably that the solutions
typically develop singularities: this already happens in the first
order Lagrangian solution \cite{Zeld70,Buch89,MABP91} (the
singularities are generically sheet-like --- `pancakes' in the
cosmological literature); the absence of any dissipative term in the
dust model Eqs.~suggests that this is a property of the model and not
an artifact of the perturbation expansion. Nevertheless, this problem
could be satisfyingly (when compared to simulations) tackled by a
phenomenological correction, the adhesion model \cite{GSS89}, which
adds a viscosity to the dust model.

Recently, I proposed a novel approach (Dom{\'\i}nguez 2000; \pI\ 
hereafter). The new model is characterized by a correction term to the
dust model featuring a length scale $L$. It relies on a formal
expansion in powers of $L$ which I call the {\em small-size expansion}
(SSE)\footnote{This anticipates the discussion in Sec.~\ref{sec:disc};
  in \pI\ the denomination was the `large-scale expansion', because it
  can be viewed also as a formal expansion in powers of $k$ in Fourier
  space.  But this designation is unfortunate, since it can be
  mistaken for the large-scale {\em cosmological} expansion.}. I
showed in \pI\ that the correction behaves like an effective viscosity
for the generic sheet-like collapse configuration; in this sense, it
represents a derivation of a generalized adhesion-like model which
continues the line of research of \cite{BuDo98,BDP99}. In that work I
also raised two questions, which will be the goal of this paper: a
better physical understanding of the length $L$ and the possibility
that the corrections may act as a source of vorticity.

The analytical approach is hindered by the nonlinear nature of the
dust model and the corrections: in Sec.~\ref{sec:pert} I apply the
Eulerian perturbation expansion in the nonlinearities, which will
allow a direct comparison to the results for the dust model obtained
with the same technique. The calculations are mathematically a bit
more involved, but there are no essential differences. The skewness of
the density contrast is computed to lowest order in
Sec.~\ref{sec:skew}, since this example provides a test case for
comparison of the SSE predictions with the dust-model ones.
Sec.~\ref{sec:vort} studies the SSE prediction to lowest
perturbational order for the vorticity of the peculiar velocity. This
is most interesting because the vorticity vanishes in the dust model
and is thus entirely due to the correction.
Sec.~\ref{sec:disc} provides a detailed discussion of the physical
meaning and interpretation of the SSE and the scale $L$.  Finally,
Sec.~\ref{sec:end} summarizes the conclusions and points out several
possible lines of future research. App.~\ref{ap:sse} briefly repeats
the derivation of the SSE in \pI. App.~\ref{ap:math} collects
mathematical manipulations.

\section{Corrections to dust and perturbation theory}
\label{sec:pert}

The model with corrections to dust which I proposed in \pI\ introduces
a comoving length scale $L$ through smoothing of the microscopic mass
and peculiar-momentum density fields. Then it is assumed that the
coupling of the small-scale ($<L$) degrees of freedom with the
large-scale ($>L$) ones is weak. This allows a certain expansion in
powers of $L$ (this is the SSE) to be truncated. App.~\ref{ap:sse}
contains a brief derivation of the model, slightly generalized over
the one in \pI\ by allowing for a time-dependent smoothing length
$L(t)$. It yields the following Eqs. in standard comoving coordinates
for the mass density field $\varrho (\bx, t)$ and the peculiar
velocity field $\bu (\bx, t)$ to first order in the SSE ($\partial_i
\equiv \partial/\partial x_i$, and a summation is implied by the
repeated index $i$):
\[
  \frac{\partial \varrho}{\partial t} = - 3 H \varrho
  - \frac{1}{a} \nabla \cdot (\varrho \bu) + 
    \frac{\dot{L}}{L} B L^2 \nabla^2 \varrho ,
\]
\[
  \frac{\partial (\varrho \bu)}{\partial t} = 
  - 4 H \varrho \bu + \varrho \bw - 
  \frac{1}{a} \nabla \cdot (\varrho \bu \bu) + \mbox{} 
\]
\[
  \quad \mbox{} + B L^2 \left\{ (\nabla \varrho \cdot \nabla) \bw - 
    \frac{1}{a} \nabla \cdot [\varrho (\partial_i \bu) 
    (\partial_i \bu)] + \frac{\dot{L}}{L} \nabla^2 ( \varrho \bu) \right\} ,
\]
\begin{equation}
  \label{eq:sse}
  \nabla \cdot \bw = - 4 \pi G a (\varrho - \varrho_b) , 
\end{equation}
\[
  \nabla \times \bw = {\bmath 0} ,
\]
where $B$ is a constant of order unity fixed by the shape of the
smoothing window. When $L$ is formally set to zero,
Eqs.~(\ref{eq:sse}) reproduce the dust model and this means physically
that the structure below the scale $L$ is dynamically unimportant. The
terms proportional to $\dot{L}$ have a simple geometrical meaning, as
discussed in App.~\ref{ap:sse}. The other two correction terms
represent the influence of the small-scale structure on the dynamical
evolution of the large scales: $(\nabla \varrho \cdot \nabla) \bw$ is
a {\em tidal} correction to the `macroscopic' gravitational field
$\bw$; $\nabla \cdot [\varrho (\partial_i \bu) (\partial_i \bu)]$ is a
correction to the advective term $\nabla \cdot (\varrho \bu \bu)$ due
to the {\em velocity dispersion} induced by shears in the
`macroscopic' velocity field $\bu$. Conceptually, these corrections
have the same origin as pressure, viscosity, heat conduction in usual
hydrodynamics, namely, the dynamical effect of the microscopic degrees
of freedom neglected in the coarsed description. But of course, their
form is very different because of the qualitatively disparate
underlying physics (long vs.~short range dominant interaction).

The nonlinear character of Eqs.~(\ref{eq:sse}) makes it difficult to
learn about the influence of the corrections without resort to some
approximation. In this work I will employ the Eulerian perturbation
expansion: introduce a bookkeeping parameter $\varepsilon$ by writing
\begin{equation}
  \label{eq:pertexp}
  \delta = \sum _{n=1}^{\infty} \varepsilon^n \delta_n, 
  \qquad \bu = \sum _{n=1}^{\infty} \varepsilon^n \bu_n.
\end{equation}
where $\delta := (\varrho/\varrho_b)-1$ is the density contrast. The
parameter $\varepsilon$ is recognized to count the degree of
nonlinearity. Perturbation theory consists in looking for a formal
solution as an expansion in powers of $\varepsilon$ which is truncated
at some order; at the end of the calculations, one sets
$\varepsilon=1$ to recover the original problem. This has been a
widely used tool to study the dust model \cite{BCGS02}.

This procedure assumes that the departures $\delta$, $\bu$ from the
homogeneous expanding background are small. In the hierarchical
scenario I will consider, this is satisfied when the fields are
observed on sufficiently large scales. This implies in turn that the
perturbational solutions should be employed in principle only to
estimate quantities which are already defined by smoothening over some
sufficiently large scale. But this is not enough, because the
nonlinearities couple widely different scales. Consequently, some of
these perturbationally computed quantities may exhibit a
short-distance, {\em ultraviolet} (UV) divergence within the dust
model ($L \rightarrow 0$); this divergence is however naturally
regularized in the context of the SSE by the finite scale $L$. This
will be seen in next Secs. with examples.

The perturbation expansions~(\ref{eq:pertexp}) are inserted in
Eqs.~(\ref{eq:sse}). The initial conditions at time $t_{in}$ read
\[
\delta_1 (\bx, t_{in}) = \delta (\bx, t_{in}) , \qquad 
\delta_\lambda (\bx, t_{in}) = 0, \quad \lambda \geq 2 ,
\]
and similarly for $\bu$. To lowest order the linearized version of
Eqs.~(\ref{eq:sse}) is found:
\[
  \frac{\partial \delta_1}{\partial t} = 
  - \frac{1}{a} \nabla \cdot \bu_1 + 
  B L \dot{L} \nabla^2 \delta_1 ,
\]
\[
  \frac{\partial \bu_1}{\partial t} = 
  - H \bu_1 + \bw_1 +
  B L \dot{L} \nabla^2 \bu_1 ,
\]
\begin{equation}
  \label{eq:first}
  \nabla \cdot \bw_1 = - 4 \pi G a \varrho_b \delta_1 , 
\end{equation}
\[
\nabla \times \bw_1 = {\bmath 0} .  
\]
These Eqs. differ from the well-known linear dust model only in the
terms $\propto \dot{L}$. The solutions are more easily written down
for the Fourier transform of the fields (to avoid burdening the
notation, the same symbol will be used for a quantity and its Fourier
transform. Which one is meant should be clear from the argument of the
function and the context),
\[
\phi(\bk) := \int {\rm d}\bx \; {\rm e}^{i \bk \cdot \bx} \; \phi(\bx).
\]
In the long-time limit ($t \gg t_{in}$, but still within the validity
regime of the perturbation expansion), the solution reads ($L_t \equiv
L(t)$, $k \equiv |\bk|$):
\begin{equation}
  \label{eq:lindens}
  \delta_1 (\bk, t) = b_t e^{-\frac{1}{2} B L_t^2 k^2} \delta_{mic} (\bk, t_{in}) ,
\end{equation}
\begin{equation}
  \label{eq:linvel}
  \bu_1 (\bk, t) = - a_t \dot{b}_t \frac{i \bk}{k^2} 
  e^{-\frac{1}{2} B L_t^2 k^2} \delta_{mic} (\bk, t_{in}) ,
\end{equation}
where $b_t$ is the growing mode of $\ddot{b} + 2 H \dot{b} - 4 \pi G
\varrho_b b =0$, with the normalization $a(t_{in})=b(t_{in})=1$, and
\[
\delta_{mic} (\bk, t_{in}) = e^{\frac{1}{2} B L_{in}^2 k^2} \delta (\bk, t_{in})
\]
is the unsmoothed initial density contrast, Eqs.~(\ref{eq:micfield}).
(This identification is rigourously true only for a Gaussian window;
otherwise, it is part of the approximation behind the SSE, see end
paragraph in App.~\ref{ap:sse}). Compared to the usual linear
solutions, Eqs.~(\ref{eq:lindens}-\ref{eq:linvel}) exhibit the
explicit time-dependent smoothing. This spoils the separability of the
time and wavevector dependences and renders the calculations a bit
more messy.

Knowing the solutions~(\ref{eq:lindens}-\ref{eq:linvel}), the
first-order nonlinear corrections obey a set of linear, inhomogeneous,
partial differential Eqs.:
\[
  \frac{\partial \delta_2}{\partial t} = 
  - \frac{1}{a} \nabla \cdot \bu_2 + 
  B L \dot{L} \nabla^2 \delta_2 - 
  \frac{1}{a} \nabla \cdot (\delta_1 \bu_1) ,
\]
\[
  \frac{\partial \bu_2}{\partial t} = 
  - H \bu_2 + \bw_2 + B L \dot{L} \nabla^2 \bu_2 - 
  \frac{1}{a} (\bu_1 \cdot \nabla) \bu_1 + \mbox{}
\]
\[
\qquad \mbox{} + B L^2 \left\{ (\nabla \delta_1 \cdot \nabla) \bw_1 - 
  {1 \over a} \nabla \cdot [(\partial_i \bu_1) (\partial_i \bu_1)] + \mbox{} \right.
\]
\begin{equation}
  \label{eq:second}
  \qquad\qquad\qquad \left. \mbox{} + 
    2 \frac{\dot{L}}{L} (\nabla \delta_1 \cdot \nabla) \bu_1 \right\} ,
\end{equation}
\[
  \nabla \cdot \bw_2 = - 4 \pi G a \varrho_b \delta_2 , 
\]
\[
  \nabla \times \bw_2 = {\bmath 0} .
\]
These Eqs. will be studied in the following Secs.
As remarked, the time-dependence of $L$ complicates a little the
calculations. However, the qualitative features of the results to
follow are the same as if $\dot{L}=0$.

\section{Skewness of the density contrast}
\label{sec:skew}

As an example of how to proceed, Eqs.~(\ref{eq:second}) are solved for
$\delta_2$ in this Sect.~and the skewness of the density contrast is
computed to leading-order. The terms $\propto \dot{L}$ are eliminated
by introducing auxiliary variables:
\[
\Delta_2 (\bk, t) := \delta_2 (\bk, t) e^{\frac{1}{2} B L_t^2 k^2} ,
\quad
{\bmath U}_2 (\bk, t) := \bu_2 (\bk, t) e^{\frac{1}{2} B L_t^2 k^2} .
\]
Eqs.~(\ref{eq:second}) then read:
\[
\frac{\partial \Delta_2}{\partial t} = \frac{1}{a} i \bk \cdot {\bmath U}_2 + 
Q_\Delta ,
\]
\begin{equation}
  \label{eq:secondnew}
  \frac{\partial {\bmath U}_2}{\partial t} = 
  - H {\bmath U}_2 - 4 \pi G a \varrho_b \frac{i \bk}{k^2} \Delta_2 + \bQ_U ,
\end{equation}
with the sources:
\[
Q_\Delta (\bk, t) := \frac{e^{\frac{1}{2} B L_t^2 k^2}}{a_t} \int \frac{{\rm d}\bq}{(2 \pi)^3} 
[ i \bk \cdot \bu_1 (\bq, t) ] \delta_1 (\bk-\bq, t) ,
\]
\[
\bQ_U (\bk, t) := e^{\frac{1}{2} B L_t^2 k^2} \int \frac{{\rm d}\bq}{(2 \pi)^3} 
\left\{ \frac{1}{a_t} [ i \bq \cdot \bu_1 (\bk-\bq, t) ] \bu_1 (\bq, t) + \mbox{} \right.
\]
\[
\quad \mbox{} + B L_t^2 [\bq \cdot (\bk-\bq)] \left[ 4 \pi G a_t
  \varrho_b \delta_1 (\bq, t) \delta_1 (\bk-\bq, t) \frac{i
    \bq}{q^2} - \mbox{} \right.
\]
\[
\qquad \left. \left. \mbox{} - \frac{1}{a_t} [i \bk \cdot \bu_1 (\bq, t)] 
\bu_1 (\bk-\bq, t) - 
\frac{2 \dot{L}_t}{L_t} \delta_1 (\bk - \bq, t) \bu_1 (\bq, t) \right] \right\} .
\]
Eqs.~(\ref{eq:secondnew}) now look the same as those of the dust model
except for the especific form of the sources. The solution for
$\Delta_2$ with the initial condition $\Delta_2 (\bk, t_{in})=0$ is
\[
  \Delta_2 (\bk, t) = b_t \int_{t_{in}}^t \!\! d\tau \int_{t_{in}}^\tau \!\! d\tau' \;
  \left(\frac{a_{\tau'} b_{\tau'}}{a_\tau b_\tau}\right)^2 
  \frac{{\cal Q} (\bk, \tau')}{b_{\tau'}} ,
\]
\[
{\cal Q} (\bk, t) := 2 H_t Q_\Delta (\bk, t) + \dot{Q}_\Delta (\bk, t) +
\frac{1}{a_t} i \bk \cdot \bQ_U (\bk, t) = \mbox{}
\]
\[
\quad \mbox = \dot{b}^2_t \int \frac{{\rm d}\bq}{(2 \pi)^3}
\delta_{mic} (\bq, t_{in}) \delta_{mic} (\bk-\bq, t_{in})
e^{\frac{1}{2} B L_t^2 (k^2 - q^2 - |\bk-\bq|^2)} \times
\]
\[
\quad \mbox{} \times \frac{\bk \cdot \bq}{q^2} [1 - B L_t^2 \; \bq \cdot (\bk - \bq)] 
  \left[ 4 \pi G \varrho_b \left(\frac{b_t}{\dot{b}_t}\right)^2 + 
    \frac{\bk \cdot (\bk-\bq)}{|\bk-\bq|^2} \right] , 
\]
where the linear solutions (\ref{eq:lindens}-\ref{eq:linvel}) have
been inserted. To simplify the solution further, we assume an
Einstein--de Sitter background, $b_t=a_t=(t/t_{in})^{2/3}$. Then,
one can write finally:
\[
\delta_2 (\bk, t) = \int \frac{{\rm d}\bq}{(2 \pi)^3} \delta_{mic}
(\bq, t_{in}) \delta_{mic} (\bk-\bq, t_{in}) \times \mbox{} 
\]
\begin{equation}
  \label{eq:pertdens}
  \qquad\qquad\qquad\qquad \mbox{} \times {\cal G}_\delta (\bk-\bq, \bq) 
  {\cal F}_\delta (\bk-\bq, \bq, t) ,
\end{equation}
with the kernel
\[
  {\cal G}_\delta (\bk, \bq) := \frac{\bq \cdot (\bk+\bq)}{q^2}
  \left[ \frac{3}{2} + \frac{\bk \cdot (\bk+\bq)}{k^2} \right] , 
\]
and the function
\[
{\cal F}_\delta (\bk, \bq, t) := a_t  e^{-\frac{1}{2} B L_t^2 |\bk+\bq|^2} 
\int_{t_{in}}^t \!\! d\tau \int_{t_{in}}^\tau \!\! d\tau' \;
  \left(\frac{a_{\tau'}}{a_\tau}\right)^4 \frac{\dot{a}^2_{\tau'}}{a_{\tau'}} \times \mbox{}
\]
\[
\qquad \mbox{} \times [1 - B L_{\tau'}^2 \, \bq \cdot \bk] 
\; e^{\frac{1}{2} B L_{\tau'}^2 (|\bk+\bq|^2 - q^2 - k^2)} .
\]
The corrections to dust are collected in the function ${\cal F}_\delta$.
In the dust limit, $L \rightarrow 0$, it becomes a function of time
only: ${\cal F}_\delta \rightarrow (2/7) a^2_t$ (in the long-time
limit $t \gg t_{in}$).

The second-order solution can be employed to estimate the skewness of
the density contrast, defined as ${\cal S}_3 := \langle \delta^3
\rangle / \langle \delta^2 \rangle^2$, where $\langle \cdots \rangle$
denotes an ensemble average over initial conditions. When the initial
density contrast is Gaussian distributed, $\langle \delta_1^3
\rangle=0$ by Eq.~(\ref{eq:lindens}). Hence, the skewness of the
evolved density contrast is generated solely by the nonlinear
evolution of the gravitational instability and this motivates the
interest in its computation.

The variance and the skewness of the density contrast smoothed on a
scale $R$ read:
\begin{equation}
  \label{eq:sigmadelta}
  \sigma^2 (R, t) = \int \frac{{\rm d}\bk \, {\rm d}\bq}{(2 \pi)^6}
  \langle \delta(\bk, t) \delta(\bq, t) \rangle
  \tilde{W} (Rk) \tilde{W} (R q) ,
\end{equation}
\[
{\cal S}_3 (R, t) = \frac{1}{\sigma^4 (R, t)} \int \frac{{\rm d}\bk \,
  {\rm d}\bq \, {\rm d}\bp}{(2 \pi)^9} \langle \delta(\bk, t)
\delta(\bq, t) \delta(\bp, t) \rangle \times
\]
\[
\qquad\qquad\qquad\qquad\qquad\qquad \times \tilde{W} (R k) \tilde{W}
(R q) \tilde{W} (R p) ,
\]
where $\tilde{W} (\cdot)$ is the Fourier transform of the smoothing
window. As mentioned previously, the use of the perturbation theory
solutions~(\ref{eq:pertexp}) requires $R$ to be large enough.
Furthermore, initial Gaussian inhomogeneities are assumed,
characterized by the power spectrum
\begin{equation}
  \label{eq:gauss}
  \langle \delta_{mic} (\bk, t_{in}) \delta_{mic} (\bq, t_{in}) \rangle = 
  (2 \pi)^3 P(k) \delta^{(3)}(\bk+\bq) .
\end{equation}
With the perturbation solutions (\ref{eq:lindens})
and~(\ref{eq:pertdens}), the variance to leading order reduces to the
linear solution,
\begin{equation}
  \label{eq:sigmalin}
  \sigma_\ell^2 (R, t) = a^2_t \int \frac{{\rm d}\bk}{(2 \pi)^3} P(k) 
   |\tilde{W} (R k)|^2 e^{-B L_t^2 k^2} ,
\end{equation}
while the skewness reads
\begin{equation}
  \label{eq:skew}
  {\cal S}_3 (R,t) = \frac{6 \, a^2_t}{\sigma_\ell^4 (R, t)} 
  \int \frac{{\rm d} \bk \, {\rm d} \bq}{(2 \pi)^6}
  P(k) P(q) {\cal G}_\delta (\bk, \bq) \times \mbox{}
\end{equation}
\[
\qquad\qquad \mbox{} \times {\cal F}_\delta (\bk, \bq, t)
\tilde{W} (R k) \tilde{W} (R q) \tilde{W} (R |\bk-\bq|) .
\]
Comparison with N-body simulations has shown that the dust prediction
($L \equiv 0$) is very good \cite{JBC93}, and not only for very large
$R$ but also even close to the nonlinear scale $r_0$, defined by the
condition $\sigma (r_0, t)=1$, where perturbation results would be
expected to break down. In fact, the limit $L \rightarrow 0$ can be
taken inside the integrals in Eqs.~(\ref{eq:sigmalin}) and
(\ref{eq:skew}), since the smoothing on scale $R$ already assures UV
convergence.
Taylor-expansion provides a correction to the dust prediction of order
$(L/R)^2$, and this suggest that the scale $L$ in the SSE is smaller
than $r_0$.

The correction to dust is barely noticeable in the leading-order
skewness because the scale $R$ already works as a short-distance
cutoff. But this is of course not the case at higher perturbational
orders or with other measurable quantities. The simplest example is
$\sigma^2 (R)$: it can be argued \cite{Vala02} that the perturbational
correction in the dust model always exhibits an UV divergence at a
sufficiently large order,
in spite of the smoothing over a scale $R$ well in the linear regime.
Indeed, the next-to-leading perturbational contribution to $\sigma^2
(R)$ (of order $\varepsilon^4$, which requires going to third order in
the perturbation expansion) exhibits the following scaling when $P(k)
\propto k^n$ (self-similar power spectrum) within the dust model
\cite{LJBH96,ScFr96b}:
\begin{equation}
  \label{eq:sigmanonlin}
  \sigma^2 (R, t) - \sigma_\ell^2 \sim \left\{
    \begin{array}[c]{l}
      \left[ {\displaystyle \frac{r_0 (t)}{R}} \right]^{2(n+3)} ,
      \quad -3 < n < - 1 , \\
      \\
      \left[ {\displaystyle \frac{r_0 (t)}{R}} \right]^{2(n+3)}
      \left[ {\displaystyle \frac{L}{R}} \right]^{-(n+1)},
      \quad - 1 < n ,
    \end{array}
  \right.
\end{equation}
with an UV cutoff $L \ll R$. If $n<-1$, the smoothening over the large
scale $R$ is enough to avoid short-distance divergences and the
scale-dependence can be guessed from self-similarity arguments.
However, if $n>-1$, there is too much small-scale power and the
non-linear couplings require an additional UV cutoff $L$.

In the next Sec., an example is studied where the deviations from the
dust model are particularly evident: the vorticity of the velocity
field.

\section{Vorticity of the peculiar velocity}
\label{sec:vort}

The dust model Eqs.~lack a source of vorticity, $\bomega := \nabla
\times \bu$:
it can arise only from an initial vorticity, which is nevertheless
strongly damped in the linear regime. This tiny initial vorticity may
be amplified in high-density, collapsing regions \cite{Buch92,BaSa93},
but then caustics develop (multistream flow, the field $\bu$ becomes
multivalued) and the dust model itself breaks down.  Thus, the
corrections to the dust model are most relevant in this case. As I
will show, Eqs.~(\ref{eq:sse}) do indeed generate vorticity, and the
leading-order perturbational contribution will be computed.

Let us take the curl of Eq.~(\ref{eq:second}) for $\bu_2$ and use that
$\bomega_1 ={\bmath 0}$ by virtue of the solution~(\ref{eq:linvel}),
to get
\begin{equation}
  \label{eq:pertomega}
  \frac{\partial \bomega_2}{\partial t} + H \bomega_2 - B L \dot{L} \nabla^2 \bomega_2 
  = \bQ_\omega ,
\end{equation}
\[
  \bQ_\omega := B L^2 \nabla \times \left\{ (\nabla \delta_1
  \cdot \nabla) \bw_1 - \frac{1}{a} \nabla \cdot [ (\partial_i
  \bu_1) (\partial_i \bu_1) ] + \mbox{} \right.
\]
\begin{equation}
  \label{eq:omegaQ}
  \qquad\qquad\qquad\qquad \left. \mbox{} + 
    2 \frac{\dot{L}}{L} (\nabla \delta_1 \cdot \nabla) \bu_1 \right\} .
\end{equation}
Upon introducing the traceless shear tensor of the peculiar velocity
field,
\[
\sigma_{ij} := \frac{1}{2} (\partial_i u_j + \partial_j u_i) -
\frac{1}{3} (\nabla \cdot \bu) \delta_{ij} ,
\]
and the tidal tensor of the peculiar gravitational
acceleration, 
\[
E_{ij} := \frac{1}{2} (\partial_i w_j + \partial_j w_i) -
\frac{1}{3} (\nabla \cdot \bw) \delta_{ij} ,
\]
one can write the source term alternatively as
\[
  \bQ_\omega = B L^2 \nabla \times \left\{ {\mathbfss E}_1 \cdot
  \nabla \delta_1 - \frac{1}{a} \nabla \cdot \left[ \frac{2}{3}
    (\nabla \cdot \bu_1) {\mathbfss \sigma}_1 + {\mathbfss
      \sigma}_1 \cdot {\mathbfss \sigma}_1 \right] + \mbox{} \right.
\]
\[
  \qquad\qquad\qquad\qquad \left. \mbox{} + 
    2 \frac{\dot{L}}{L} (\nabla \delta_1 \cdot \nabla) \bu_1 \right\} .
\]
Since ${\mathbfss \sigma}_1$ is in fact proportional to ${\mathbfss
  E}_1$ by Eq.~(\ref{eq:linvel}), the ultimate dynamical source of
vorticity is the gravitational tidal force, whose effect is manifested
directly (first term) and indirectly through the induced velocity
dispersion (second and third terms). As already remarked, the fourth
term has a purely geometrical origin.

Inserting the linear solutions (\ref{eq:lindens}-\ref{eq:linvel}) in
Eq.~(\ref{eq:omegaQ}) and Fourier transforming, this source reads
\[
\bQ_\omega (\bk, t) =  B L^2 \frac{{\cal F}_\omega(t)}{a_t} 
\int \frac{{\rm d} \bq}{(2 \pi)^3} \delta_{mic} (\bq, t_{in}) 
\delta_{mic} (\bk-\bq, t_{in}) \times \mbox{} 
\]
\begin{equation}
  \label{eq:omsource}
  \qquad\qquad\qquad \mbox{} \times \frac{(\bk \cdot \bq) (\bk
    \times \bq)}{q^2} \, e^{-\frac{1}{2} B L_t^2 (q^2+|\bk-\bq|^2)} .
\end{equation}
\[
{\cal F}_\omega (t) := (a \dot{b})^2 \left[ 1 + 4 \pi G \varrho_b
  \left(\frac{b}{\dot{b}}\right)^2 + 2 \frac{b \dot{L}}{\dot{b} L}
\right] .
\]
Eq.~(\ref{eq:pertomega}) can be integrated immediately for the Fourier
transform of the vorticity with the initial condition $\bomega_2 (\bk,
t_{in})={\bmath 0}$:
\begin{equation}
  \label{eq:omega2}
  \bomega_2 (\bk, t) = \int_{t_{in}}^t \!\! d\tau \; \frac{a_\tau}{a_t} \,
  e^{-\frac{1}{2} B (L_t^2-L_\tau^2) k^2} \bQ_\omega (\bk, \tau) .  
\end{equation}

An estimate of the strength of the vorticity on a scale $R$ is
provided by its variance, 
\[
\sigma^2_\omega (R, t) := \int \frac{{\rm d}\bk \, {\rm d} \bq}{(2
  \pi)^6} \langle \bomega(\bk, t) \cdot \bomega(\bq, t) \rangle
\tilde{W} (Rk) \tilde{W} (R q) .
\]
Inserting the perturbation
solution~(\ref{eq:omsource}-\ref{eq:omega2}) and simplifying:
\[
  \sigma^2_\omega (R, t) = \int \frac{{\rm d} \bk \, {\rm d} \bq}{(2 \pi)^6}
  P(k) P(q) {\cal G}_\omega (\bk, \bq)  |\tilde{W} (R |\bk-\bq|)|^2 \times \mbox{}
\]
\begin{equation}
  \label{eq:varomega} 
  \qquad\qquad\qquad \mbox{} \times |\Xi (\bk \cdot \bq, t)|^2 
  \, e^{-B L_t^2 |\bk - \bq|^2} ,
\end{equation}
with the positive, symmetrized kernel
\[
{\cal G}_\omega (\bk, \bq) := \frac{1}{2} |\bk \times \bq|^2 (\bk \cdot \bq)^2
\left[ \frac{1}{k^2} - \frac{1}{q^2} \right]^2 ,
\]
and the function
\begin{equation}
  \label{eq:xi}
  \Xi (z,t) := \frac{1}{a_t} \int_{t_{in}}^t \!\! d\tau \;
  B L_\tau^2 {\cal F}_\omega(\tau) \, e^{-B L_\tau^2 z} .
\end{equation}

To gain insight into this result, it will be worked out for the
particular case of a self-similar model: an Einstein--de Sitter
background and an initial power spectrum $P(k) \propto k^n$ are
assumed. It remains to specify the smoothing length $L(t)$. In a
self-similar model, the only relevant scale is $r_0(t)$ and the
natural choice is $L(t) = \lambda r_0(t)$, where $\lambda$ is a
time-independent proportionality constant. This choice will be
supported by the discussion of the physical meaning of the SSE in the
next Sec.

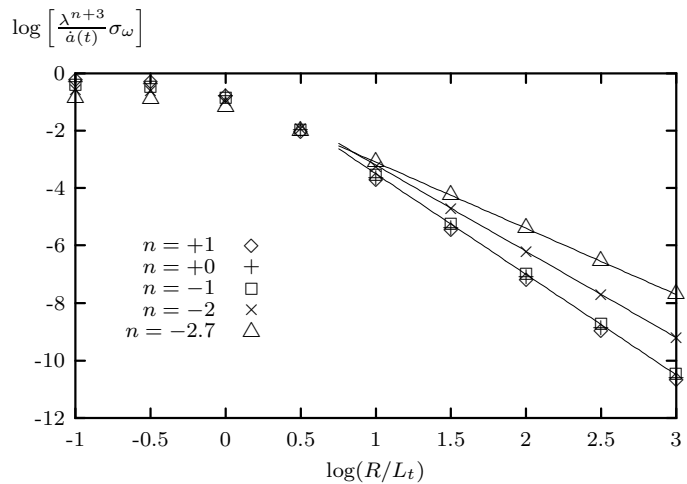
\begin{figure}
  \begin{center}
    \bigskip

\setlength{\unitlength}{0.180900pt}
\begin{picture}(1500,900)(100,0)
\footnotesize
\thicklines \path(177,135)(197,135)
\thicklines \path(1433,135)(1413,135)
\put(155,135){\makebox(0,0)[r]{-12}}
\thicklines \path(177,255)(197,255)
\thicklines \path(1433,255)(1413,255)
\put(155,255){\makebox(0,0)[r]{-10}}
\thicklines \path(177,375)(197,375)
\thicklines \path(1433,375)(1413,375)
\put(155,375){\makebox(0,0)[r]{-8}}
\thicklines \path(177,496)(197,496)
\thicklines \path(1433,496)(1413,496)
\put(155,496){\makebox(0,0)[r]{-6}}
\thicklines \path(177,616)(197,616)
\thicklines \path(1433,616)(1413,616)
\put(155,616){\makebox(0,0)[r]{-4}}
\thicklines \path(177,736)(197,736)
\thicklines \path(1433,736)(1413,736)
\put(155,736){\makebox(0,0)[r]{-2}}
\thicklines \path(177,856)(197,856)
\thicklines \path(1433,856)(1413,856)
\put(155,856){\makebox(0,0)[r]{0}}
\thicklines \path(177,135)(177,155)
\thicklines \path(177,856)(177,836)
\put(177,90){\makebox(0,0){-1}}
\thicklines \path(334,135)(334,155)
\thicklines \path(334,856)(334,836)
\put(334,90){\makebox(0,0){-0.5}}
\thicklines \path(491,135)(491,155)
\thicklines \path(491,856)(491,836)
\put(491,90){\makebox(0,0){0}}
\thicklines \path(648,135)(648,155)
\thicklines \path(648,856)(648,836)
\put(648,90){\makebox(0,0){0.5}}
\thicklines \path(805,135)(805,155)
\thicklines \path(805,856)(805,836)
\put(805,90){\makebox(0,0){1}}
\thicklines \path(962,135)(962,155)
\thicklines \path(962,856)(962,836)
\put(962,90){\makebox(0,0){1.5}}
\thicklines \path(1119,135)(1119,155)
\thicklines \path(1119,856)(1119,836)
\put(1119,90){\makebox(0,0){2}}
\thicklines \path(1276,135)(1276,155)
\thicklines \path(1276,856)(1276,836)
\put(1276,90){\makebox(0,0){2.5}}
\thicklines \path(1433,135)(1433,155)
\thicklines \path(1433,856)(1433,836)
\put(1433,90){\makebox(0,0){3}}
\thicklines \path(177,135)(1433,135)(1433,856)(177,856)(177,135)
\put(44,950){\makebox(0,0)[l]{\shortstack{$\log \left[ \frac{\lambda^{n+3}}{\dot{a}(t)} \sigma_\omega \right]$}}}
\put(805,23){\makebox(0,0){$\log (R/L_t)$}}
\put(469,496){\makebox(0,0)[r]{$n=+1$}}
\put(177,844){\raisebox{-1.2pt}{\makebox(0,0){$\Diamond$}}}
\put(334,839){\raisebox{-1.2pt}{\makebox(0,0){$\Diamond$}}}
\put(491,811){\raisebox{-1.2pt}{\makebox(0,0){$\Diamond$}}}
\put(648,736){\raisebox{-1.2pt}{\makebox(0,0){$\Diamond$}}}
\put(805,635){\raisebox{-1.2pt}{\makebox(0,0){$\Diamond$}}}
\put(962,530){\raisebox{-1.2pt}{\makebox(0,0){$\Diamond$}}}
\put(1119,425){\raisebox{-1.2pt}{\makebox(0,0){$\Diamond$}}}
\put(1276,320){\raisebox{-1.2pt}{\makebox(0,0){$\Diamond$}}}
\put(1433,215){\raisebox{-1.2pt}{\makebox(0,0){$\Diamond$}}}
\put(545,496){\raisebox{-1.2pt}{\makebox(0,0){$\Diamond$}}}
\put(469,451){\makebox(0,0)[r]{$n=+0$}}
\put(177,839){\makebox(0,0){$+$}}
\put(334,835){\makebox(0,0){$+$}}
\put(491,809){\makebox(0,0){$+$}}
\put(648,737){\makebox(0,0){$+$}}
\put(805,638){\makebox(0,0){$+$}}
\put(962,534){\makebox(0,0){$+$}}
\put(1119,429){\makebox(0,0){$+$}}
\put(1276,324){\makebox(0,0){$+$}}
\put(1433,219){\makebox(0,0){$+$}}
\put(545,451){\makebox(0,0){$+$}}
\put(469,406){\makebox(0,0)[r]{$n=-1$}}
\put(177,833){\raisebox{-1.2pt}{\makebox(0,0){$\Box$}}}
\put(334,829){\raisebox{-1.2pt}{\makebox(0,0){$\Box$}}}
\put(491,806){\raisebox{-1.2pt}{\makebox(0,0){$\Box$}}}
\put(648,739){\raisebox{-1.2pt}{\makebox(0,0){$\Box$}}}
\put(805,645){\raisebox{-1.2pt}{\makebox(0,0){$\Box$}}}
\put(962,543){\raisebox{-1.2pt}{\makebox(0,0){$\Box$}}}
\put(1119,439){\raisebox{-1.2pt}{\makebox(0,0){$\Box$}}}
\put(1276,334){\raisebox{-1.2pt}{\makebox(0,0){$\Box$}}}
\put(1433,229){\raisebox{-1.2pt}{\makebox(0,0){$\Box$}}}
\put(545,406){\raisebox{-1.2pt}{\makebox(0,0){$\Box$}}}
\put(469,361){\makebox(0,0)[r]{$n=-2$}}
\put(177,822){\makebox(0,0){$\times$}}
\put(334,819){\makebox(0,0){$\times$}}
\put(491,799){\makebox(0,0){$\times$}}
\put(648,741){\makebox(0,0){$\times$}}
\put(805,660){\makebox(0,0){$\times$}}
\put(962,572){\makebox(0,0){$\times$}}
\put(1119,483){\makebox(0,0){$\times$}}
\put(1276,393){\makebox(0,0){$\times$}}
\put(1433,303){\makebox(0,0){$\times$}}
\put(545,361){\makebox(0,0){$\times$}}
\put(469,316){\makebox(0,0)[r]{$n=-2.7$}}
\put(177,805){\makebox(0,0){$\triangle$}}
\put(334,803){\makebox(0,0){$\triangle$}}
\put(491,786){\makebox(0,0){$\triangle$}}
\put(648,736){\makebox(0,0){$\triangle$}}
\put(805,671){\makebox(0,0){$\triangle$}}
\put(962,602){\makebox(0,0){$\triangle$}}
\put(1119,533){\makebox(0,0){$\triangle$}}
\put(1276,464){\makebox(0,0){$\triangle$}}
\put(1433,395){\makebox(0,0){$\triangle$}}
\put(545,316){\makebox(0,0){$\triangle$}}
\thinlines \path(727,698)(727,698)(734,694)(741,689)(748,684)(755,679)(762,674)(769,670)(776,665)(784,660)(791,655)(798,650)(805,646)(812,641)(819,636)(826,631)(834,627)(841,622)(848,617)(855,612)(862,607)(869,603)(876,598)(884,593)(891,588)(898,584)(905,579)(912,574)(919,569)(926,564)(933,560)(941,555)(948,550)(955,545)(962,541)(969,536)(976,531)(983,526)(991,521)(998,517)(1005,512)(1012,507)(1019,502)(1026,498)(1033,493)(1041,488)(1048,483)(1055,478)(1062,474)(1069,469)(1076,464)
\thinlines \path(1076,464)(1083,459)(1090,455)(1098,450)(1105,445)(1112,440)(1119,435)(1126,431)(1133,426)(1140,421)(1148,416)(1155,412)(1162,407)(1169,402)(1176,397)(1183,392)(1190,388)(1198,383)(1205,378)(1212,373)(1219,369)(1226,364)(1233,359)(1240,354)(1247,349)(1255,345)(1262,340)(1269,335)(1276,330)(1283,325)(1290,321)(1297,316)(1305,311)(1312,306)(1319,302)(1326,297)(1333,292)(1340,287)(1347,282)(1355,278)(1362,273)(1369,268)(1376,263)(1383,259)(1390,254)(1397,249)(1404,244)(1412,239)(1419,235)(1426,230)(1433,225)
\thinlines \path(727,709)(727,709)(734,705)(741,701)(748,697)(755,692)(762,688)(769,684)(776,680)(784,676)(791,672)(798,668)(805,664)(812,660)(819,656)(826,651)(834,647)(841,643)(848,639)(855,635)(862,631)(869,627)(876,623)(884,619)(891,615)(898,610)(905,606)(912,602)(919,598)(926,594)(933,590)(941,586)(948,582)(955,578)(962,574)(969,570)(976,565)(983,561)(991,557)(998,553)(1005,549)(1012,545)(1019,541)(1026,537)(1033,533)(1041,529)(1048,524)(1055,520)(1062,516)(1069,512)(1076,508)
\thinlines \path(1076,508)(1083,504)(1090,500)(1098,496)(1105,492)(1112,488)(1119,483)(1126,479)(1133,475)(1140,471)(1148,467)(1155,463)(1162,459)(1169,455)(1176,451)(1183,447)(1190,443)(1198,438)(1205,434)(1212,430)(1219,426)(1226,422)(1233,418)(1240,414)(1247,410)(1255,406)(1262,402)(1269,397)(1276,393)(1283,389)(1290,385)(1297,381)(1305,377)(1312,373)(1319,369)(1326,365)(1333,361)(1340,356)(1347,352)(1355,348)(1362,344)(1369,340)(1376,336)(1383,332)(1390,328)(1397,324)(1404,320)(1412,316)(1419,311)(1426,307)(1433,303)
\thinlines \path(727,704)(727,704)(734,701)(741,698)(748,695)(755,692)(762,689)(769,685)(776,682)(784,679)(791,676)(798,673)(805,670)(812,667)(819,663)(826,660)(834,657)(841,654)(848,651)(855,648)(862,645)(869,641)(876,638)(884,635)(891,632)(898,629)(905,626)(912,623)(919,619)(926,616)(933,613)(941,610)(948,607)(955,604)(962,601)(969,598)(976,594)(983,591)(991,588)(998,585)(1005,582)(1012,579)(1019,576)(1026,572)(1033,569)(1041,566)(1048,563)(1055,560)(1062,557)(1069,554)(1076,550)
\thinlines \path(1076,550)(1083,547)(1090,544)(1098,541)(1105,538)(1112,535)(1119,532)(1126,528)(1133,525)(1140,522)(1148,519)(1155,516)(1162,513)(1169,510)(1176,506)(1183,503)(1190,500)(1198,497)(1205,494)(1212,491)(1219,488)(1226,484)(1233,481)(1240,478)(1247,475)(1255,472)(1262,469)(1269,466)(1276,462)(1283,459)(1290,456)(1297,453)(1305,450)(1312,447)(1319,444)(1326,440)(1333,437)(1340,434)(1347,431)(1355,428)(1362,425)(1369,422)(1376,418)(1383,415)(1390,412)(1397,409)(1404,406)(1412,403)(1419,400)(1426,396)(1433,393)
\end{picture}
    
    \caption{      
      Log-log plot of the numerically computed $\sigma_\omega$ as a
      function of $R/L_t$ in the long-time limit for an initial $P(k)
      \propto k^n$ and for a Gaussian smoothing window. The lines have
      the slope predicted by the asymptotic scaling
      (\ref{eq:scaling}).  }
    \label{fig:sigmaomega}
  \end{center}
\end{figure}

The details of the mathematical analysis are collected in
App.~\ref{ap:math}. The physically interesting results can be
summarized in the asymptotic behavior of $\sigma_\omega$ in the limit
$R \gg L_t$ and $t \gg t_{in}$:
\begin{equation}
  \label{eq:scaling}
  \frac{\sigma_\omega (R, t)}{\dot{a} (t)} \sim \left\{
  \begin{array}[c]{l}
    \left[ {\displaystyle \frac{L_t}{R}} \right]^{2} 
    \left[ {\displaystyle \frac{r_0 (t)}{R}} \right]^{n+3} ,
    \quad -3 < n < - {\displaystyle \frac{3}{2}} , \\
    \\
    \left[ {\displaystyle \frac{L_t}{R}} \right]^{2} 
    \left[ {\displaystyle \frac{r_0 (t)}{R}} \right]^{n+3}
    \left[ {\displaystyle \frac{L_t}{R}} \right]^{-n-\frac{3}{2}},
    \quad -{\displaystyle \frac{3}{2}} < n ,
  \end{array}
  \right.
\end{equation}
up to a dimensionless factor which depends only on the spectral index
$n$. The lower bound $-3<n$ prevents an {\em infrared} (IR)
divergence. The crossover at $n=-3/2$ is related to the UV behavior of
the integrals, and we are in a similar situation as with
Eq.~(\ref{eq:sigmanonlin}). If $n<-3/2$, $R$ is enough to regularize
possible UV divergences, the scale $L$ only appears in the prefactor
$B L^2$ of the source (\ref{eq:omegaQ}) and the scale-dependence of
$\sigma_\omega$ can be guessed by simple power-counting. If $n>-3/2$,
however, $L$ is relevant also as UV cutoff. Fig.~\ref{fig:sigmaomega}
shows the numerical computation of $\sigma_\omega$ with
Eq.~(\ref{eq:varomega}), confirming the asymptotic scaling
(\ref{eq:scaling}); in particular, that the $R$-dependence is
insensitive to the spectral index when $n>-3/2$. It also shows how
well the asymptotic scaling is already obeyed when $R/L_t \sim 10$. In
the opposite limit $R \rightarrow 0$ (known in the literature as the
`unsmoothed' limit), $\sigma_\omega$ approaches a nonvanishing
constant.

To address the relevance of the generated vorticity, it will be
compared to the dust prediction for the divergence of the velocity
field, $\theta := \nabla \cdot \bu$. The variance $\sigma_\theta^2 :=
\langle \theta^2 \rangle$ on a scale $R$ is defined in analogy with
Eq.~(\ref{eq:sigmadelta}). 
The variance $\sigma^2_\omega$, of perturbational order
$\varepsilon^4$, must be compared with the next-to-leading order in
$\sigma^2_\theta$, denoted by $\tilde{\sigma}^2_\theta$. This term
scales like $\dot{a}^2$ times Eq.~(\ref{eq:sigmanonlin}); more
generally, each term in the perturbational expansion of
$\sigma^2_\theta$ is proportional to the corresponding one in the
perturbational expansion of $(\dot{a} \sigma)^2$
\cite{ScFr96a,FoGa98}. Thus
\begin{equation}
  \label{eq:ratio}
  \frac{\sigma_\omega}{\tilde{\sigma}_\theta} \sim \left\{
    \begin{array}[c]{ll}
      \left[ {\displaystyle \frac{L_t}{R}} \right]^{2} , &
      \quad -3 < n < - {\displaystyle \frac{3}{2}} , \\
      & \\
      \left[ {\displaystyle \frac{L_t}{R}} \right]^{-n+\frac{1}{2}} , &
      \quad -{\displaystyle \frac{3}{2}} < n < -1 , \\
      & \\
      \left[ {\displaystyle \frac{L_t}{R}} \right]^{-\frac{n}{2}+1} , &
      \quad -1 < n ,
    \end{array}
  \right.
\end{equation}
up to a factor of order unity. The exponent of the ratio $L_t/R$ is in
all cases positive. At fixed resolution $R$, the vorticity becomes
more and more relevant in time; at fixed time, it becomes subdominant
on large scales.

A more realistic CDM initial power spectrum at the scales of physical
interest can be approximated well by a power-law $P(k) \propto k^n$
with $n \approx -1, -2$. This initial condition is set at a redshift
$z \sim 10^3$, and so the present epoch corresponds to $a_t \sim
10^3$, well in the long-time regime (see App.~\ref{ap:math}).
Realistic deviations from an exact Einstein--de Sitter background are
expected to affect only weakly the results, as happens with
predictions by the dust model. As will be argued in
Sec.~\ref{sec:disc}, the ratio $\lambda=L_t/r_0(t)$ is $\sim 0.1$.
With a value $r_0 (\textrm{today}) \approx 8\ h^{-1}$Mpc,
the estimate~(\ref{eq:ratio}) suggests that the vorticity becomes
significant today on scales $R \approx L_t \approx 1\ h^{-1}$Mpc; the
physical vorticity ${\bmath \omega} / a$ has then a value of the order
of $1$-$10$ times the Hubble constant today, as can be immediately
read from Fig.~\ref{fig:sigmaomega}. This conclusion does not disprove
the approximation of potential flow at weakly nonlinear, large scales
($R \gg r_0(t)$), which is invariably done both in theoretical
\cite{SaCo95} and observational \cite{Deke94} studies. Both the length
scale of relevance and the magnitude of the vorticity agree with the
result by Pichon \& Bernardeau \shortcite{PiBe99}, who studied the
growth of vorticity with the Zel'dovich aproximation at times just
after the formation of the first caustic. But the SSE offers a more
straightforward and systematic method than the one employed by these
authors.

\section{Discussion}
\label{sec:disc}

As follows from the derivation in App.~\ref{ap:sse}, the SSE relies on
a dynamically negligible effect of the substructure of the coarsening
cells of size $\approx L$; the only dynamically relevant properties of
the cells are the mass, $\propto \varrho$, and the center-of-mass
velocity, $\bu$. On the other hand, one may notice that
Eqs.~(\ref{eq:newton}-\ref{eq:poisson}) for the evolution of {\em
  point particles} under its own gravity can be formally written as
the dust model Eqs. for the fields $\varrho_{mic}$ and $\bu_{mic}$,
Eqs.~(\ref{eq:micfield}). Therefore, an interpretation of the
assumption behind the SSE is that the evolution of the scales $>L$
corresponds, in the lowest approximation, to that of a set of
`effective point particles', plus corrections due to the nonvanishing
particle size $\approx L$ and their internal structure.  This
interpretation motivates the appellation `small-size expansion'.  In a
bottom-up scenario, it is natural to identify the effective particles
with the most recently (and thus largest) formed clusters.  In fact,
Peebles (1974; 1980, Sec.~28) had already shown the mutual
cancellation of the nonlinear couplings to the small scales in the
Eq.~for $\delta$ when these small scales consist of virialized
clusters.

It must be noticed that this provides a different physical
interpretation of the dust model and extends its validity. Indeed, the
dust model is usually justified starting from the ideal fluid model:
then the pressure term is dropped because self-gravity is dominant on
scales much larger than the Jeans' length. This argumentation applies
in top-down ('pancake') models because, at a given time, matter is
smoothly distributed on small scales and the free-fly of particles
gives rise to a kinetic pressure: the ideal fluid model is a good
approximation. This does not hold, however, in the opposite limit of a
bottom-up ('hierarchical') evolution, for the small scales are highly
structured. Nevertheless, the SSE argues how the dust model can still
be a good description when the evolution is observed through the right
spatial resolution $\sim L^{-1}$: then particles appear trapped in the
virialized clusters and to a first approximation, there is no
free-fly, no kinetic pressure. In both extreme scenarios is the dust
model the lowest-order approximation, the differences show up only in
the corrections.

The Zel'dovich approximation \cite{Zeld70} was originally devised to
address structure formation in top-down scenarios (HDM models). It is
the lowest-order term in a Lagrangian perturbation expansion of the
dust model \cite{Buch89,MABP91}. Thus, it came out as unexpected and
worth remarking that it provided also a good description of the
large-scale structure in bottom-up simulations, particularly if the
initial power spectrum was first smoothed on a scale $\sim r_0$: this
is the {\em truncated} Zel'dovich approximation
\cite{CMS93,MPS94,PaMe95}. The SSE offers a physical explanation for
this convergence and, as a novel aspect, implements in a systematic
way the role of the `truncation' (smoothing) and the `residual'
coupling to the neglected small scales.

We see that smoothing is an essential component of the dust model
itself in a hierarchical scenario. The use of an UV-cutoff in the
perturbation expansion is not new, but it was mainly introduced as a
matter of practical purposes; e.g., to regularize the divergences in
the perturbation expansion of the dust model as the initial
small-scale power is increased (Sec.~\ref{sec:skew}).  These
divergences are regarded unphysical, since N-body simulations show
nothing special in the observables. This is usually interpreted as a
failure of the perturbation expansion, assuming the results from
simulations to be qualitatively right, and so it is conjectured that
divergences could be cured by resumming the expansion \cite{Bern96}.
The SSE can be viewed as such a resummation: it takes account of a
highly nonlinear feature --- the virialized clusters, and its
dynamical effect on the evolution of the large scales.

According to the physical interpretation of the SSE, the smoothing
length $L(t)$ should scale with $r_0(t)$, as the natural measure of
the size of the largest virialized structures. Consequently, the
self-similar behavior expected when $P(k) \propto k^n$ on an
Einstein--de Sitter background is not spoiled. Such a cutoff was in
fact employed by Jain \& Bertschinger \shortcite{JaBe96}, and
criticized in turn by Scoccimarro \& Frieman \shortcite{ScFr96b}, who
rejected its {\it ad hoc} time dependence. The latter authors suggest
that self-similarity breaks down in general in perturbation theory,
because the UV cutoff is just a property of the initial conditions.
The SSE approach reconciles self-similarity with perturbation theory
by realizing that the dynamically generated UV cutoff $L(t) \propto
r_0 (t)$ is itself a {\em defining} ingredient of the dust model in a
hierarchical scenario.

Structures of size $L(t)$ should behave approximately as `particles'
(clusters); hence, $L(t)$ must be defined through a condition which
explicitly tests somehow the degree of `dynamical isolation' of the
structure below a given scale. In Sec.~\ref{sec:skew} it was concluded
that $L$ is different from the nonlinear scale $r_0$: then, $\sigma^2
= 1$ is not such a tester. Nevertheless, a possible suggestion of how
the function $\sigma(R,t)$ could be useful is the following: Let us
assume that, in the case of a scale-invariant initial power spectrum
and an Einstein--de Sitter background, $\sigma(R, t)$ is given by the
stable clustering ansatz \cite{DaPe77} in the nonlinear regime ($R \ll
r_0$).  Then, the recently formed clusters, and thus the length $L$,
may be identified by the scale $R$ at which $\sigma(R,t)$ starts
following the stable clustering prediction. Fig.~3 in \cite{CBH96} and
Fig.~1 in \cite{Jain97} provide the estimate $L/r_0 \sim 10^{-1}$.
N-body simulations seem to support the stable clustering hypothesis
\cite{EFWD88,CBH96,Jain97,VLS00}. But this point is not settled yet
and there are evidences that it only holds for the recently formed
clusters \cite{MCCM97,MaFr00}: even so, however, the previous
reasoning would still be functional.
The numerical results for the vorticity, Sec.~\ref{sec:vort}, also
favor a ratio $L/r_0 \sim 0.1$.

This estimate is consistent with the fact that the dust model
predictions agree with N-body simulations {\em when the former are not
  UV-divergent}, in some cases also up to the limit $R \approx r_0$
(beyond the {\it a priori} expected range of validity, $R \gg r_0$):
as the example of the skewness in Sec.~\ref{sec:skew} shows,
corrections are then of order $(L/r_0)^2$ when pushing $R \approx
r_0$, and thus possibly unobservable (e.g., the estimated absolute
error of the measured skewness is about $0.1$ \cite{JBC93}).  On the
one hand, this good agreement of the convergent dust-model predictions
supports the hypothesis of the mode-decoupling at the basis of the
SSE. On the other hand, it suggests that even the dust-model
UV-divergent perturbation results may provide good approximations when
regularized in the framework of the SSE.

\section{Conclusions and Outlook}
\label{sec:end}

I have explored the application of Eulerian perturbation theory to a
recently proposed model of cosmological structure formation, the
small-size expansion (SSE), which incorporates corrections to the dust
model. The SSE features a length scale $L(t)$ which is identified with
the typical size of the most recently formed virialized structures.
The perturbation results suggest $L/r_0 \sim 0.1$ with $r_0$ the
nonlinear scale.

When the dust-model perturbation results are finite, the corrections
are irrelevant. However, the length $L(t)$ works as a natural
UV-cutoff which regularizes perturbation expansions in the dust model
which would be otherwise divergent. The SSE may also address
multistreaming effects perturbationally, which are absent in the dust
model. As an example, I have computed the vorticity generated by tidal
torques: its relative importance is predicted to diminish when the
cosmic flow is observed on increasingly larger scales, and to be
significant today on scales $\sim 1\ h^{-1}$Mpc with a value $\sim$
Hubble constant.

The SSE provides a new perspective and a possibly useful tool in the
analytical understanding of the evolution by gravitational
instability. The model based on the SSE must still be certainly
subject to further investigation to check its internal consistency and
eventual success: the results of this work can be contrasted to N-body
simulations. As well as the predictions for the vorticity, it will be
particularly interesting to compare regularized perturbation
corrections, e.g., that for the density variance $\sigma^2$. Finally,
another interesting line of future research on the SSE is the
application of the {\it Lagrangian} perturbation theory: this should
help clarify the relationship to the truncated Zel'dovich
approximation.

\section*{Acknowledgments}

The author acknowledges C.~Beisbart, T.~Buchert, J.~Gaite,
M.~Kerscher, and the anonymous referee for their useful remarks on the
manuscript. Most part of this work was completed at the Department for
Theoretical Physics in the Ludwig-Maximilians-Universit\"at, Munich.

\appendix

\section{Derivation of the small-size expansion (SSE)}
\label{ap:sse}

To make this work as self-contained as possible, I provide in this
App. the derivation in \pI\ leading to Eqs.~(\ref{eq:sse}). It is
however more general in that a time-dependent smoothing length is
allowed.

The basic model is a system of nonrelativistic, identical point
particles which (i) are assumed to interact with each other via
gravity only, (ii) look homogeneously distributed on sufficiently
large scales, which thus are assumed to evolve according to an
expanding Friedmann-Lema\^\i tre cosmological background, and (iii)
deviations to homogeneity are relevant only on scales small enough
that a Newtonian approximation is valid to follow their evolution. Let
$a(t)$ denote the expansion factor of the Friedmann-Lema\^\i tre
cosmological background, $H(t) = \dot{a}/a$ the associated Hubble
function, and $\varrho_b (t)$ the homogeneous (background) density on
large scales.  $\bx_\alpha$ is the comoving spatial coordinate of the
$\alpha$-th particle, $\bu_\alpha$ is its peculiar velocity, and $m$
its mass. In terms of these variables the evolution is described by
the following set of equations ($\nabla_\alpha$ denotes a partial
derivative with respect to $\bx_\alpha$):
\begin{equation}
  \label{eq:newton}
  \dot{\bx}_\alpha = {1 \over a} \bu_\alpha , \qquad
  \dot{\bu}_\alpha = \bw_\alpha - H \bu_\alpha ,
\end{equation}
\begin{equation}
  \label{eq:poisson}
  \nabla_\alpha \cdot \bw_\alpha = - 4 \pi G a \left[ {m \over a^3} 
    \sum_{\beta \neq \alpha} \delta(\bx_\alpha - \bx_\beta) - \varrho_b \right] , 
\end{equation}
\[
  \nabla_\alpha \times \bw_\alpha = {\bmath 0} ,
\]
where $\bw_\alpha$ is the peculiar gravitational acceleration acting on
the $\alpha$-th particle. Finally, Eqs. (\ref{eq:poisson}) must be subjected
to periodic boundary conditions in order to yield a Newtonian
description consistent with the Friedmann-Lema\^\i tre solution at
large scales \cite{EhBu97}.

One can formally define a microscopic mass density field and a
microscopic peculiar velocity field,
\[
\varrho_{mic} (\bx, t) = \frac{m}{a(t)^3} \sum_\alpha \delta^{(3)}
(\bx - \bx_\alpha (t)) ,
\]
\begin{equation}
  \label{eq:micfield}
  \varrho_{mic} \bu_{mic} (\bx, t) = \frac{m}{a(t)^3} \sum_\alpha
  \bu_\alpha (t) \delta^{(3)} (\bx - \bx_\alpha (t)) .
\end{equation}
Coarsening cells of comoving size $\sim L(t)$ are defined with the
help of a smoothing window $W(z)$ (see \pI\ for details on the
properties required on the window). The coarse-grained fields
associated to the microscopic fields are respectively
\[
\varrho (\bx, t) = \int {{\rm d}\by \over L^3} \; W \left( {|\bx -
    \by| \over L} \right) \varrho_{mic} (\by, t) ,
\]
\begin{equation}
  \label{eq:smoothing}
  \varrho \bu (\bx, t) = \int {{\rm d}\by \over L^3} \; W \left( {|\bx
      - \by| \over L} \right) \varrho_{mic} \bu_{mic} (\by, t) .
\end{equation}
Notice in particular that $\bu$ is defined by a volume-average of the
peculiar momentum density (i.e., by a {\em mass-average} of the
velocity) and thus its physical meaning is transparent: $\bu$ is the
center-of-mass velocity of the coarsening cell. These fields obey the
following exact evolution Eqs. (conservation of mass and momentum,
following from Eqs.~(\ref{eq:newton})):
\[
  \frac{\partial \varrho}{\partial t} = - 3 H \varrho
  - \frac{1}{a} \nabla \cdot (\varrho \bu) 
  + \frac{\dot{L}}{L} \nabla \cdot {\bmath s} ,
\]
\begin{equation}
  \label{eq:exact}
  \frac{\partial (\varrho \bu)}{\partial t} = 
  - 4 H \varrho \bu + {\bmath f} - 
  \frac{1}{a} \nabla \cdot {\mathbfss P}
    + \frac{\dot{L}}{L} \nabla \cdot {\mathbfss D} .
\end{equation}
New fields have arisen, whose physical meaning is also clear from its
definitions: two vector fields,
\[
{\bmath s} (\bx, t) = \int {{\rm d}\by \over L^3} \; W \left( {|\bx
    - \by| \over L} \right) (\by - \bx) \varrho_{mic} (\by, t) ,
\]
\[
{\bmath f} (\bx, t) = \int {{\rm d}\by \over L^3} \; W \left( {|\bx
    - \by| \over L} \right) \varrho_{mic} \bw_{mic} (\by, t) ,
\]
and two second-rank tensor fields,
\begin{equation}
  \label{eq:veldisp}
  {\mathbfss P} (\bx, t) = \int {{\rm d}\by \over L^3} \; 
  W \left( {|\bx - \by| \over L} \right) 
  \varrho_{mic} \bu_{mic} \bu_{mic}
  (\by, t) ,
\end{equation}
\[
  {\mathbfss D} (\bx, t) = \int {{\rm d}\by \over L^3} \; 
  W \left( {|\bx - \by| \over L} \right)  (\by - \bx)
  \varrho_{mic} \bu_{mic} (\by, t) .
\]
(The definition of $\bw_{mic}$ is Eq.~(\ref{eq:micfield}) with the
replacement $\bu_\alpha \rightarrow \bw_\alpha$). 
If the evolution Eqs. for these fields are computed, new fields
appear, and so on {\it ad infinitum}. A physical assumption is
required to cut the hierarchy: the nonlinear coupling in
Eqs.~(\ref{eq:exact}) to the degrees of freedom below the smoothing
scale $L$ is assumed to be weak. To implement this idea
mathematically, notice that Eqs.~(\ref{eq:smoothing}) can be formally
inverted to yield the following expansion in $L$ (this is most easily
derived in Fourier space):
\[
\varrho_{mic} = \left[ 1 - {1 \over 2} B \, (L \, \nabla)^2 + o(L
\, \nabla)^4 \right] \, \varrho ,
\]
\begin{equation}
  \label{eq:micexpan}
  \bu_{mic} = \left[ 1 - {1 \over 2} B \, (L \, \nabla)^2 - B L^2
    {\nabla \varrho \over \varrho} \cdot \nabla + o(L \, \nabla)^4
  \right] \, \bu ,
\end{equation}
with the constant
\[
B = {1 \over 3} \int {\rm d}{\bmath z} \; z^2 \, W(z) = {4 \pi \over 3}
\int_0^{+\infty} {\rm d} z \; z^4 \, W(z) .
\]
These expansions just show how the microscopic fields can be recovered
by taking into account higher and higher derivatives of the smoothed
fields, i.e., finer and finer details in the spatial distribution. Now
consider, e.g., the tensor field ${\mathbfss P}$, encompassing
nonlinear mode-mode couplings: the weak-coupling assumption above then
means that the dominant dynamical contribution of ${\mathbfss P}$ to
the evolution Eq.~(\ref{eq:exact}) arises from the coupling between
modes on scales $>L$. Hence, ${\mathbfss P}$ can be replaced in this
Eq. by the result of inserting the expansions~(\ref{eq:micexpan}) in
the definition~(\ref{eq:veldisp}):
\[
{\mathbfss P} \rightarrow \varrho \bu \bu + B L^2 \varrho (\partial_i
\bu) (\partial_i \bu) + o(L \, \nabla)^4 .
\]
Applying the same reasoning to the other fields:
\[
{\bmath f} \rightarrow \varrho \bw + B L^2 (\nabla \varrho \cdot \nabla)
\bw + o(L \, \nabla)^4 ,
\]
\[
  \nabla \cdot \bw = - 4 \pi G a (\varrho - \varrho_b) , 
\]
\[
  \nabla \times \bw = {\bmath 0} ,
\]
and
\[
{\bmath s} \rightarrow B L^2 \nabla \varrho + o(L \, \nabla)^4 ,
\]
\begin{equation}
  \label{eq:diff}
  {\mathbfss D} \rightarrow B L^2 \nabla (\varrho \bu) + o(L \, \nabla)^4 .
\end{equation}
Taking these results in Eqs.~(\ref{eq:exact}), one obtains a formal
expansion in $L$ (called the small-size expansion in the main text).
Truncating after the second term, Eqs.~(\ref{eq:sse}) are recovered.

The fields ${\bmath s}$ and ${\mathbfss D}$ have a purely geometrical
origin and describe how the smoothed fields change simply because the
smoothing length is varied. Its diffusion-like expression is thus
barely surprising. In fact, truncating the expansions~(\ref{eq:diff})
is {\em exact} for a Gaussian window, $W(z) = \exp (-\pi z^2)$ and $B=1/2 \pi$. In
general, the approximation~(\ref{eq:diff}) is equivalent to replacing
the arbitrary window $W(z)$ by a Gaussian one with a width fixed by
the constant $B$.

\section{Vorticity variance} 
\label{ap:math}

In this App.~I collect the mathematical details of the study of
Eq.~(\ref{eq:varomega}). 

We consider first the IR behavior. When $P(k) \sim k^n$ as $k
\rightarrow 0$, the integrals behave asymptotically in the limit $q
\rightarrow 0$, $(k/q) \rightarrow 0$ as ($s := \cos \phi$, $\phi :=
\widehat{(\bk, \bq)}$)
\[
\sim |\Xi (0, t)|^2 \int_0 \!\! {\rm d} k \; k^{n+2} \int_0 \!\! {\rm
  d} q \; q^{n+6} \int_{-1}^{+1} \!\!\!\! {\rm d} s \; s^2 (1-s^2) ,
\]
which is finite and nonvanishing if $n>-3$, independently of the
values of $R$ and $L$.

To address the UV behavior ($k, q \rightarrow +\infty$), one makes the
change of variable $\bq \rightarrow \bp=\bk-\bq$.
Assuming $P(k) \sim k^n$ as $k \rightarrow +\infty$, we have
asymptotically
\[
\sim \int^{+\infty} \!\! {\rm d} k \; k^{2n+2} |\Xi (k^2, t)|^2
\int_0^{+\infty} \!\!\!\! {\rm d} p \; p^{4} e^{-B L_t^2 p^2} |\tilde{W} (R p)|^2 \times
\]
\[
\;\;\times \int_{-1}^{+1} \!\!\!\! {\rm d} s \; s^2 (1-s^2) .
\]
The smoothing window over scale $R$ renders the $p$-integral
UV-convergent, no matter the value of $L$. From
the definition of the function $\Xi$, it follows that $\Xi (k^2, t)
\leq \Xi (0, t) e^{-B L_{in}^2 k^2}$ (since $\dot{L} > 0$ by
assumption), so that the $k$-integral is also convergent as long as $L_{in}
\neq 0$, as expected. However, in the dust limit $L \equiv 0$, the
$k$-integral converges only if $n<-3/2$.

In what follows, the particular case of an Einstein--de Sitter
background is considered with the self-similar initial power spectrum
\begin{equation}
  \label{eq:pk}
  P(k) = A_n r_{in}^{n+3} k^n, \qquad 
  \frac{1}{A_n} := \int \frac{{\rm d} \bk}{(2 \pi)^3} k^n |\tilde{W}(k)|^2 \; , 
\end{equation}
where $r_{in} \equiv r_0 (t_{in})$. A Gaussian window is assumed,
$W(z) = \exp (-\pi z^2)$, and then
\[
A_n = \frac{(2 \pi)^{\frac{1-n}{2}}}{\Gamma \left( \frac{n+3}{2}
  \right)} .
\]
In Eq.~(\ref{eq:varomega}), three angular integrations can be
performed immediately. Then, one exploits the symmetry of the
integrand under exchange $\bk \leftrightarrow \bq$ and changes to new
variables $k \rightarrow k/L_t$, $q \rightarrow k p$:
\[
\sigma_\omega^2 = \frac{2 A_n^2}{(2 \pi)^4} \frac{r_{in}^{2(n+3)}}{L_t^{2(n+5)}} 
\int_0^{+\infty} \!\!\!\! {\rm d} k \; k^{2n+9}
\int_0^{1} \!\!\!\! {\rm d} p \; p^{n+2} (1-p^2)^2 \times \mbox{}
\]
\[
\quad \mbox{} \times 
 \int_{-1}^{+1} \!\!\!\! {\rm d} s \; s^2 (1-s^2)
 \left| \Xi \left(\frac{k^2 p s}{L_t^2}, t \right) \right|^2 \,  
 e^{-B \left\{1+\left[\frac{R}{L_t}\right]^2 \right\} (1+p^2-2 p s) k^2} .
\]
Inserting Eq.~(\ref{eq:pk}) in Eq.~(\ref{eq:sigmalin}), one finds the
linear dust prediction (i.e., up to corrections of order $(L/R)^2$):
$r_0 (t) = r_{in} a^{2/(n+3)}_t$. As remarked in Sec.~\ref{sec:vort}, a
constant ratio $\lambda = L_t / r_0(t)$ is taken. Then, the function
$\Xi$
can be written in terms of the incomplete gamma function. However, it
proves a better idea (particularly in order to compute $\sigma_\omega$
numerically) to perform first the $k$-integral, and next, in the
integral defining $\Xi$, Eq.~(\ref{eq:xi}), to change variables $\tau
\rightarrow [r_0(t)/r_{in}] \tau = a_t^{4/(n+3)} \tau$. One finally
gets:
\[
\sigma_\omega^2 = \frac{\Gamma(n+5)}{2^{n+3}}
\left[ \frac{(5n+23)  \dot{a}_t}{8 \Gamma \left(\frac{n+3}{2}\right)} \right]^2 
\left[ \frac{r_0(t)}{L_t} \right]^{2(n+3)} 
\left[1+\left(\frac{R}{L_t}\right)^2 \right]^{-(n+5)}
\]
\begin{equation}
  \label{eq:integral}
  \quad \mbox{} \times
  \int^1_{(r_{in}/r_0)^2} \!\! {\rm d} \tau_1 \, {\rm d} \tau_2 \; 
  (\tau_1 \tau_2)^{\frac{5(n+3)}{8}} \, 
  H_n \left(\frac{\tau_1+\tau_2}{1+(R/L_t)^2} \right) , \qquad
\end{equation}
with the function
\[
H_n (y) := \int_0^1 \!\!\!\! {\rm d} z \; z^{n+2} \sqrt{1-z^2} 
\int_{-1}^{+1} \!\!\!\! {\rm d} s \; s^2 (1-s^2) 
\left( 1 - \frac{2-y}{2} z s \right)^{-(n+5)} ,
\]
obtained with another change of variable: $z := 2 p/ (1+p^2)$. Only
the domain $0 \leq y \leq 2$ is interesting. The function may exhibit
a singularity as $y \rightarrow 0^+$. A straightforward study of the
convergence of the integrals near the limits $s=1$, $z=1$ provides the
asymptotic behavior
\[
H_n (y \rightarrow 0^+ ) \sim \left\{ 
    \begin{array}[c]{ll}
      1 , & -3 < n < - {\displaystyle \frac{3}{2}} , \\
      & \\
      y^{-n-\frac{3}{2}} , & - {\displaystyle \frac{3}{2}} < n . \\
    \end{array}
    \right.
\]
With this information, one can easily derive the scaling behavior
(\ref{eq:scaling}) from Eq.~(\ref{eq:integral}). The long-time limit,
$t \gg t_{in}$, is obtained by setting the lower limit of the
$\tau$-integrals to zero (no divergence arises with the constraint
$-3<n$). In the exponent range $-3<n<4$, the long-time approximation
is accurate to the $1\%$ level for times such that $a_t > 5$.

The integral (\ref{eq:integral}) is also well suited for numerical
evaluation. In fact, the $s$-integral in the function $H_n$ can be
computed analytically. And the introduction of the variable $u :=
\tau_1+\tau_2$
allows to compute one of the $\tau$-integrals analytically too. In the
end, the numerical computation reduces to a two-dimensional integral
in the finite domain $0 \leq z \leq 1$, $0 \leq u \leq 2$.

\end{document}